# Scalable Deterministic Task Offloading and Resource Allocation in the IoT-Edge-Cloud Continuum

Keyvan Aghababaiyan, Baldomero Coll-Perales, Javier Gozalvez
*Uwicore laboratory, Universidad Miguel Hernandez de Elche, Elche (Alicante), Spain*
kaghababaiyan@umh.es, bcoll@umh.es, j.gozalvez@umh.es

*Abstract*— Future 6G networks are envisioned as a "network of networks" (NoN) ecosystem, integrating communication and computing resources across multiple domains. At the deep edge, IoT and end-user devices will form subnetworks for local communication and distributed task processing. These subnetworks will seamlessly integrate into the NoN ecosystem, creating an IoT-edge-cloud continuum. The unified resources across this continuum facilitate dynamic and scalable task offloading, unlocking new possibilities to support emerging services, including critical vertical services with stringent reliability and deterministic service level requirements. In this context, this paper demonstrates that a deterministic approach to task offloading and resource (communication and computing) allocation in the IoT-edge-cloud continuum not only ensures deterministic service levels but also enhances scalability compared to existing task offloading and resource allocation methods. By flexibly managing task completion deadlines while maintaining deterministic (i.e. bounded latency) service levels, deterministic policies achieve a more balanced workload and resource distribution across the continuum, ultimately improving scalability.

## I. Introduction

Beyond 5G (B5G) and 6G networks are envisioned as a 'network of networks' (NoN) ecosystem, integrating diverse communication networks to enable seamless and ubiquitous connectivity [1]. This includes subnetworks deployed at the deep edge of the network for local communications (see [1], [2], and 3GPP TSG-SA WG1, e.g., S1-240121, S1-244238). Subnetworks are composed of different types of IoT devices (e.g. sensors and actuators) that can seamlessly interconnect either locally – via sidelink or direct communications- or with a wide-area cellular network for providing cost-effective service delivery for applications with diverse requirements while supporting distributed processing for autonomous local data management. The B5G and 6G vision goes beyond pure communication systems, aiming to sustainably integrate computing, communication and intelligence into a unified system capable of supporting the ever-growing demands across an IoT-edge-cloud continuum [3][4]. This continuum provides a programmable computing infrastructure across IoT devices, edge and cloud nodes that expands capabilities and flexibility for dynamically deploying applications and network services while adapting deployments to variable demands.

This work has been partially funded by the European Commission Horizon Europe SNS JU 6G-SHINE (GA 101095738) project, and by MCIN/AEI/10.13039/501100011033 (PID2020-115576RB-I00, PID2023-150308OB-I00).

Realizing the potential of the IoT-edge-cloud continuum requires efficient task offloading and resource allocation strategies to dynamically distribute tasks across resources in the continuum. These strategies should consider the requirements from various types of applications including emerging ones like critical vertical applications in fields such as industrial automation, cyber-physical systems, healthcare, or autonomous mobility. Many of these applications demand high dependability and deterministic service levels, which existing networks cannot easily provide. However, they can benefit from the distributed computing resources within the IoT-edge-cloud continuum. In fact, deterministic communications and networking have been identified as key enablers in 6G to support emerging and critical vertical applications at scale [5]. Future networks and systems must efficiently scale while maintaining the required dependability and deterministic service levels, even as the number of connected devices, computational demands, spectrum constraints and stringent communication requirements continue to grow. Achieving this scalability in a 6G-based NoN ecosystem operating an IoT-edge-cloud continuum depends on the effectiveness of task offloading and resource allocation strategies in efficiently managing communication and computing resources in the continuum.

This paper advances the state of the art by demonstrating that a deterministic approach to task offloading and resource allocation not only ensures the required deterministic service levels but also scales more effectively than existing task offloading and resource allocation strategies. To this end, we compare the scalability of a deterministic policy with a state-of-the-art policy that seeks minimizing task execution time. Our results show that, by flexibly managing task completion deadlines, a deterministic approach to task offloading and resource allocation achieves a more balanced workload and resource distribution across the continuum. This, in turn, improves the system's ability to meet task execution deadlines for a larger number of tasks and nodes, enhancing overall scalability. Our evaluation further demonstrates that a deterministic policy leads to higher task completion rates, improved fairness across the system, and greater adaptability to variations in computing and communication resource utilization and conditions.

## II. Architecture and System Modeling

The 6G NoN vision integrates multiple networks and domains in a unified framework. This includes subnetworks for local IoT communications, e.g. at factories for communications between robots or collaborative robots (cobots), as well as within and between autonomous vehicles

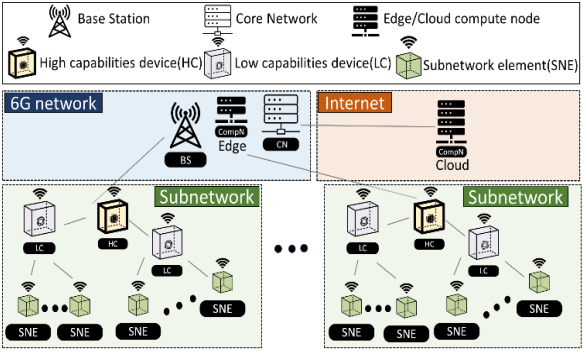

Fig. 1. IoT-edge-cloud architecture.

[1][2]. This integration provides the possibility to establish an IoT-edge-cloud continuum where tasks can be seamlessly offloaded across the continuum. Fig. 1 shows the IoT-edge-cloud architecture that integrates subnetworks for local connectivity at the deep edge. Each subnetwork comprises three main components: Subnetwork Elements (SNEs), Low Capability (LC) units, and a High Capability (HC) unit [2]. The HC unit is equipped with advanced computational and networking capabilities, serving as the central hub of the subnetwork. It can handle intensive tasks and connect to the 6G wide-area network to access resources at the edge and the cloud. LC units, with more limited capabilities compared to HC, handle less computationally intensive tasks and serve as intermediaries between the SNEs and the HC unit. The SNEs are resource-constrained devices, like sensors and actuators, which may require support from LC and HC for the processing of the tasks they generate.

In this IoT-edge-cloud architecture, we consider a collection of tasks $f_{i,n}$ ($i \in \{1, ..., I\}$) generated within the subnetwork $n$ ($n \in \{1, ..., N\}$) at time instant $t_{i,n}$. Tasks can be generated by $SNE_n$, $LC_n$, and $HC_n$ of the subnetwork $n$. Each task $f_{i,n}$ is defined by its computing demand $c_{i,n}$ and associated size $s_{i,n}$. When a task is offloaded to a processing unit different from where it was generated, the processed result, which has a reduced size $s'_{i,n}$ compared to the original size $s_{i,n}$, must be transmitted back to its source unit. Each task has a deadline $T_{i,n}^{max}$, which indicates the maximum time available to complete processing. For tasks handled by a different unit, the deadline accounts for the transmission time to move the task to the processing unit, the processing duration, and the time required to transmit back the processed result. Computing units have different processing capacities, denoted by $P_x$, where $x$ refers to the type of processing unit. $x_s \in \{LC_n, HC_n\}$ represents local processors of the subnetworks, while $x_p \in \{Ed, Cl\}$ refers to the edge ($Ed$) and cloud ($Cl$) units. The time required to process a task $f_{i,n}$ on a computing unit $x \in \{x_s, x_p\}$ is given by:

$$t_p^{i,n} = \frac{c_{i,n}}{P_x}. \quad (1)$$

We consider that wireless communication links within the subnetworks as well as between the subnetworks and the wide-area cellular network utilize an Orthogonal Frequency Division Multiple Access (OFDMA)-based radio access interface. Subnetworks have a dedicated communication band with a bandwidth of $BW_s$ that does not overlap with the band used for communication with the wide-area cellular network, which has a bandwidth $BW_p$. Inter-subnetwork interference is not considered based on recent subnetwork channel characterization measurements [6] that demonstrate the possibility to isolate subnetworks through well-planned and characterized environments (e.g. introducing directive communications and reconfigurable intelligence surfaces), and the high penetration losses of material within the subnetworks (e.g., within vehicles). The subnetworks' bandwidth $BW_s$ is divided into $K_s$ orthogonal communication resources, which can be reused within different subnetworks. $BW_p$ is divided into $K_p$ orthogonal resources, and these resources are shared among the $N$ subnetworks for their connectivity with the wide-area cellular network. The data rate available at any given time for communication resource $k \in \{K_s, K_p\}$ in link $l \in \{1, 2, .., L\}$, either within the subnetwork $n$ or from the subnetwork $n$ to the wide-area network, is denoted as $r_{l,n}^{(k)}(t)$ and can be expressed as:

$$r_{l,n}^{(k)}(t) = BW_k \cdot \log_2\left(1 + \gamma_{l,n}(t)\right)(1 - BER). \quad (2)$$

In (2), $BW_k$ represents the bandwidth of the communication resource $k$, $\gamma_{l,n}(t)$ denotes the Signal-to-Interference plus Noise Ratio (SINR) at time $t$ of the link $l$, and BER is the bit error rate, which depends on the modulation and coding scheme employed in the communication resource $k$. To model channel fading effects, we assume a Rayleigh distribution. The total data rate of a communication link $l$ is calculated as the sum of the data rates for all communication resources $k$ utilized in the link:

$$r_{l,n}(t) = \sum_k r_{l,n}^{(k)}(t). \quad (3)$$

When a task $f_{i,n}$ requires offloading, the transmission time over different communication links $l_i \in \{1, 2, .., L\}$ is determined as:

$$t_c^{i,n} = \sum_{l_i} \frac{s_{i,n}}{r_{l_i,n}(t)}. \quad (4)$$

Similarly, the transmission time over communication links $l_i \in \{1, 2, .., L\}$ for the processed result of a task $f_{i,n}$ with size of $s'_{i,n}$ can be expressed as $t'^{i,n}_c$ and is computed following (4) using $s'_{i,n}$ instead of $s_{i,n}$. The total time $T_{i,n}$ required to execute a task $f_{i,n}$ generated within the subnetwork $n$, includes the communication time for moving the task from its source to the processing unit ($t_c^{i,n}$), the processing time at the computing unit ($t_p^{i,n}$), and the communication time for returning the processed result ($t'^{i,n}_c$):

$$T_{i,n} = t_c^{i,n} + t_p^{i,n} + t'^{i,n}_c. \quad (5)$$

III. TASK OFFLOADING AND RESOURCE ALLOCATION

Realizing the potential of the IoT-edge-cloud continuum requires task offloading and resource allocation strategies that jointly manage and optimize communication and computing resources across the continuum [4]. Several strategies have been proposed to date, with a common primary focus for most of them on minimizing computational and communication latencies, which is particularly relevant for low latency applications. For example, Cai et al. [7] analyzed the trade-offs between local and remote task processing to minimize latency while considering bandwidth and processing power constraints. In [8], the authors propose a reverse offloading framework to reduce system latency by opportunistically utilizing resources either at the edge or at vehicles to process large amounts of data. Similarly, the proposal in [9] seeks to minimize processing delays by leveraging idle resources at devices and offloading tasks to these devices under increasing processing demands. In [10], Oliveira et al. propose a task allocation strategy that minimizes response times for latency-sensitive applications while reducing network traffic by mitigating idle resource time in hierarchical fog architectures. In contrast to minimizing latency or task execution time, we advocate for a deterministic approach to task offloading and resource allocation that prioritizes increasing the number of tasks executed within their deadlines over reducing task execution or completion time. Deterministic schemes can flexibly manage task completion deadlines to balance workload and resource distribution across the continuum, which in turn has the potential to enhance scalability by improving the system's ability to meet task execution deadlines for a larger number of tasks. To evaluate this potential, we compare the scalability of a deterministic task offloading and resource allocation scheme (referred to as *Deterministic* in this study) against a reference state-of-the-art strategy focused on reducing task completion time. Specifically, we compare the performance of a deterministic scheme against a reference scheme from [7] referred to as *Minimum* in this paper, that we implement in our system model presented in Section II along with a random strategy. The following sub-sections describe the objective functions and common system constraints for these three strategies.

### A. Objective functions

*Minimum* allocates communication and computing resources within the IoT-edge-cloud continuum with the primary objective of minimizing task execution time. Its objective function can be formulated as:

$$min \sum_n \sum_i T_{i,n}, \qquad (6)$$

where $T_{i,n}$, as defined in (5), represents the execution time of the task $f_{i,n}$ generated in the subnetwork $n$.

The *Deterministic* scheme is designed to ensure that tasks are executed before their deadlines (i.e. $T_{i,n} < T_{i,n}^{max}$) rather than focusing on minimizing execution time. The objective is for *Deterministic* to leverage the flexibility and varying deadlines of tasks to distribute and balance them across the IoT-edge-cloud continuum. Its objective function is formulated as:

$$min \sum_n \sum_i \beta\left(\frac{T_{i,n}}{T_{i,n}^{max}}\right), \qquad (7)$$

where $T_{i,n}$ represents the execution time of task $f_{i,n}$ generated in the subnetwork $n$, $T_{i,n}^{max}$ is the task's deadline, and $\beta$ is a penalty function defined as:

$$\beta(\xi) = \begin{cases} 0, & 0 \leq \xi \leq 1, \\ M, & \xi \geq 1. \end{cases} \qquad (8)$$

In (8), $\xi$ represents the normalized execution time $T_{i,n}$ of a task relative to its deadline $T_{i,n}^{max}$, and $M$ is a large positive constant. The penalty function in (8) is introduced to ensure a deterministic execution of tasks within their bounded deadlines, as it penalizes tasks whose execution $T_{i,n}$ exceeds the deadline $T_{i,n}^{max}$.

A baseline third *Random* scheme randomly selects the computing unit to process the task $T_{i,n}$ generated in the subnetwork $n$. If the selected unit requires moving the task to another unit of the IoT-edge-cloud continuum, the communication resources needed to transmit the task to the selected unit are also chosen randomly from the available options. Nevertheless, the communication and computing resources are allocated to ensure that each task is processed within its deadline $T_{i,n}^{max}$.

### B. Common constraints to all considered schemes

For fairness, all schemes (*Minimum*, *Deterministic* and *Random*) are defined with six common constraints in the task offloading and resource allocation processes.

The first constraint is the binary task allocation constraint, which states that task $f_{i,n}$ generated in subnetwork $n$ is allocated to a single computing unit and cannot be divided among multiple units. The mathematical expression for this constraint is given as:

$$\sum_{j=1}^{J} a_{i,n}^{(j)} + \sum_{q=1}^{Q} a_{i,n}^{(q)} + a_{i,n}^{(c)} = 1, \quad \forall\, i, n, \qquad (9)$$

where $J$ is the number of local computing units in the subnetwork $n$ (i.e., the sum of $LC_n$, and $HC_n$ units) and $Q$ is the number of edge computing units. $a_{i,n}^{(j)}$, $a_{i,n}^{(q)}$ and $a_{i,n}^{(c)}$ are binary variables equal to 1 if task $f_{i,n}$ generated in subnetwork $n$ is allocated to the local computing unit $j$, the edge unit $q$, or the cloud unit $c$, respectively, and 0 otherwise.

In accordance with the principles of OFDMA, the second constraint ensures that communication resources assigned to subnetworks (i.e., $k_s \in K_s$) can be utilized by only one communication link $l$ at a time, preventing transmissions collisions within the subnetwork. The mathematical formulation of this constraint is expressed as:

$$\sum_l \sum_i b_{l,i,n}^{(k_s)} \leq 1, \qquad \forall\, k_s, n, \qquad (10)$$

where $b_{l,i,n}^{(k_s)}$ is a binary variable equal to 1 when communication resource $k_s$ is allocated to transmit task $f_{i,n}$ generated in the subnetwork $n$ to the link $l$.

The third constraint ensures that the shared communication resources for subnetworks to connect to the wide-area cellular network (i.e., $k_p \in K_p$) can be allocated to only one communication link at a time. Hence, this constraint prevents transmissions collisions between subnetworks accessing the wide-area cellular network. The mathematical expression for this constraint is given as:

$$\sum_n \sum_l \sum_i b_{l,i,n}^{(k_p)} \leq 1, \qquad \forall\, k_p, \qquad (11)$$

where $b_{l,i,n}^{(k_p)}$ is a binary variable equal to 1 when communication resource $k_p$ is allocated to transmit task $f_{i,n}$ in link $l$ between the subnetwork $n$ and the wide-area network.

The fourth constraint ensures that the total transmission rate of all tasks utilizing links between subnetworks and the wide-area network does not exceed the maximum achievable data rate of that link. The mathematical expression of this constraint is:

$$\sum_i r_{l,i,n}(t) \leq r_{l,n}(t), \quad \forall l, n, \tag{12}$$

where $r_{l,i,n}(t)$ is the data rate of link $l$ for transmitting task $f_{i,n}$ generated in subnetwork $n$ and $r_{l,n}(t)$ is the maximum possible data rate of link $l$.

The fifth and sixth constraints ensure that the total processing workload of tasks assigned to a local computing unit within subnetwork $n$ (eq. (13)) or to an edge and cloud computing units (eq. (14)) does not exceed the maximum processing capacity of that unit over a given time interval. The mathematical formulation of these constraints are given as:

$$\sum_i c_{i,n} a_{i,n}^{(x_s)} \leq C_{x_s,n}^{max}, \quad \forall x_s, n, \tag{13}$$

$$\sum_i \sum_n c_{i,n} a_{i,n}^{(x_2)} \leq C_{x_p}^{max}, \quad \forall x_p, \tag{14}$$

where $C_{x_s,n}^{max}$ is the maximum processing capacity of computing unit $x_s$ in subnetwork $n$, and $C_{x_p}^{max}$ is the maximum processing capacity of computing unit $x_p$ (edge or cloud nodes).

## IV. EVALUATION

### A. Scenario

Without loss of generality, we evaluate the scalability of the task offloading and resource allocation schemes in a 6G-envisioned industrial cyber-physical scenario in which wireless subnetworks formed by mobile robots are connected with a cellular network [11]. This scenario enables the processing of the tasks generated by the robots in the IoT-edge-cloud continuum. We evaluate scenarios involving varying numbers of subnetworks ($N$), ranging from 2 to 5, and different numbers of tasks ($I$), ranging from 5 to 25. We consider each subnetwork includes 15 SNEs representing robots' sensors, and 4 LCs and 1 HC computing units for processing, monitoring, and control tasks within the subnetwork. Subnetwork components can be placed anywhere within a subnetwork and do not move between subnetworks during the simulation period. While tasks can be generated by any element in the subnetwork, local processing within subnetworks is limited to LCs and HCs. The processing power of {LC, HC, edge, and cloud} is {2.5, 5, 70 and 150} GHz following [12]. Within each subnetwork, {60%, 20%, and 20%} of tasks are generated by {SNEs, LCs, and HC}, respectively. We consider tasks processing workloads and sizes to emulate the varying requirements of traffic generated from mobile robots. Following [13], we consider that mobile robots tasks require between 20 and 50 Mcycles and range in size from 0.75 to 2.25 Mbits, with an average size of 1 Mbits. The processed result size for each task is set to 15% of its original size. Following the requirements for cyber-physical control of mobile robots defined in [14], task deadlines ($T_i^{max}$) are randomly allocated within the 20 to 100 ms range. The penalty value $M$ in (8) assigned to tasks that are not completed by their deadlines is 100. Subnetworks are configured to operate with a dedicated bandwidth $BW_s$ of 100 MHz, while the links between subnetworks and the wide-area cellular network use a bandwidth $BW_p$ of 50 MHz. Following [6], we assume that wireless links within subnetworks maintain an average SINR of 30 dB, with the channel fading modelled with a Rayleigh distribution. The average SINR for the links between subnetworks and the wide-area cellular networks varies between 0 dB and 30 dB. We consider a subcarrier spacing (SCS) of 30 kHz and a time slot duration of 0.5 ms [15].

We have implemented a genetic algorithm in MATLAB to solve the optimization problems of the task offloading and resource allocation schemes. Considering the number of variables and possible resource allocation options in the evaluated scenarios, the optimization process starts with a population of a thousand candidate resource allocation solutions. The top 20% of the best-performing candidates are retained for the next generation, while the remaining 80% are created through crossover operations from the top 20%. The algorithm iterates over ten generations. A mutation rate of 20% introduces small random changes to enhance diversity and prevent premature convergence. The configuration of the algorithm balances performance and computation complexity, and our tests showed that it converged to near-optimal solutions [16]. The deterministic and benchmark schemes are compared for the same number of generations, ensuring a fair comparison by maintaining identical run times.

### B. Results

Fig. 2 and Fig. 3 depict the average ratio of satisfied tasks as a function of the number of executed tasks per subnetwork and number of subnetworks, under good and poor SINR conditions in the connection to the wide-area cellular network, respectively. The satisfaction ratio represents the proportion of tasks completed before their deadlines relative to the total number of tasks and is averaged across all subnetworks. The obtained results show that *Deterministic* achieves the highest satisfaction ratio, regardless of the number of tasks, the number of subnetworks, or the connection quality to the wide-area network. For example, the obtained results show that the *Deterministic* scheme supports all 25 tasks generated across the 5 different subnetworks, with each subnetwork generating 5 tasks. In contrast, the number of satisfied tasks is reduced by an average of {10%, 38%} for SINR=30dB (Fig. 2) and {18%, 45%} for SINR=0dB (Fig. 3) for the *Minimum* and *Random* schemes, respectively. Increasing the number of tasks per subnetwork reduces the satisfaction ratio for all schemes due to the limited computing and communication resources available in the scenario compared to the simulated workloads. However, the reduction observed is less significant for the *Deterministic* scheme, which highlights its better scalability. This is also observed as the number of subnetworks augments. In this case, the satisfaction ratio does not decrease with the *Deterministic*, while the satisfaction ratio for *Minimum* and *Random* decreases as the number of subnetworks increase. These results emphasize the scalability advantage of deterministic policies that jointly allocate and manage communication and computing resources across the IoT-edge-cloud continuum with the objective of satisfying the tasks' execution deadlines rather than minimizing tasks' execution time. Trying to minimize each task's execution time can result in many tasks trying to use computing units with higher computing capabilities or links with high data rates, increasing the pressure on these resources. The results obtained show that this can ultimately lead to the overload of these resources and a lower satisfaction ratio and poorer scalability compared to deterministic policies. This is particularly visible under bad link quality conditions (Fig. 3) as *Minimum* can result in satisfaction ratios even lower than with *Random* for the highest number of subnetworks and of tasks per subnetwork analyzed. The *Deterministic* scheme focuses on ensuring all tasks are executed before their deadlines rather than minimizing their execution time. *Deterministic* leverages the flexibility and varying deadlines of tasks to distribute and balance tasks across the continuum.

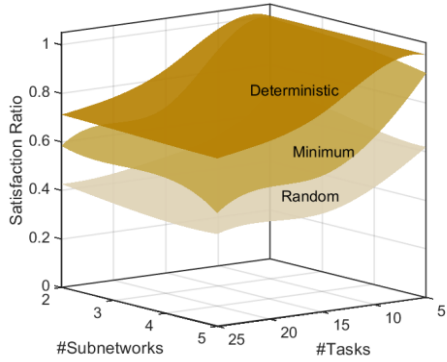

Fig. 2. Average ratio of satisfied tasks as a function of number of subnetworks and tasks per subnetworks when average SINR=30 dB.

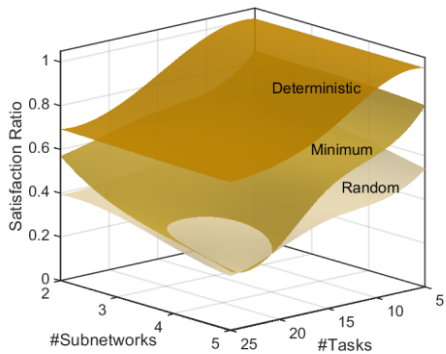

Fig. 3. Average ratio of satisfied tasks as a function of number of subnetworks and tasks per subnetworks when average SINR=0 dB.

By balancing the workload across the continuum, *Deterministic* avoids putting excessive strain on certain computing and communication resources, thus improving

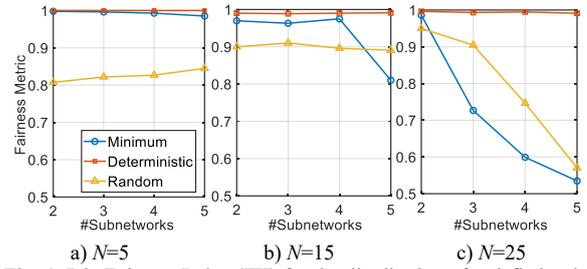

Fig. 4. Jain Fairness Index (JFI) for the distribution of satisfied tasks for average SINR = 30 dB (similar trends observed for SINR = 0 dB).

overall scalability.

The scalability of the task allocation schemes also depends on the fairness in task satisfaction across subnetworks. The average satisfaction ratio shown in Fig. 2 and Fig. 3 does not reflect the balance in task satisfaction across subnetworks. Fig. 4 reports the Jain Fairness Index (JFI) metric to measure the distribution of task satisfaction ratios among different subnetworks [17]. The JFI for satisfaction ratio of a set of subnetworks $(SR_n, n \, \epsilon \{1, \dots, N\})$ can be expressed as:

$$JFI(SR_n) = \frac{(\sum_n SR_n)^2}{N \sum_n SR_n^2}. \quad (15)$$

The JFI takes values ranging from 0 to 1. A value close to 1 signifies a more equitable distribution of the satisfaction ratio among different subnetworks, while a value closer to 0 indicates significant disparities. Fig. 4 reports the measured JFI for scenarios with 2 to 5 subnetworks and an increasing number of tasks per subnetwork: 5 in Fig. 4.a, 15 in Fig. 4.b and 25 in Fig. 4.c. The obtained results demonstrate that *Deterministic* consistently achieves a JFI value close to 1 across all scenarios, regardless of the number of tasks or subnetworks. This indicates a high level of fairness in the distribution of satisfied tasks among different subnetworks. In contrast, the JFI values for *Minimum* and *Random* decrease as the number of subnetworks and tasks increases, which highlights lower fairness and significant variations in the number of satisfied tasks across subnetworks.

The scalability benefits of *Deterministic* over the other evaluated schemes stem from its better distribution and balance of tasks and workload across the IoT-edge-cloud continuum, achieved by leveraging the varying deadlines of tasks. This is illustrated in Fig. 5, which depicts the average ratio of utilized communication and computing resources of the link(s) and processing unit selected by the task offloading and resource allocation schemes[1] relative to the total number of resources. The utilization ratio is measured from the moment the task is allocated to the selected link(s) or processing unit until its deadline. The bars indicate the average resource utilization ratio, while the lines within each bar represent the standard deviation. The obtained results show that *Deterministic* selects less congested communication

---

[1]. The previous analysis has shown that *Random* is outperformed by the other schemes and is omitted from the figure for clarity.

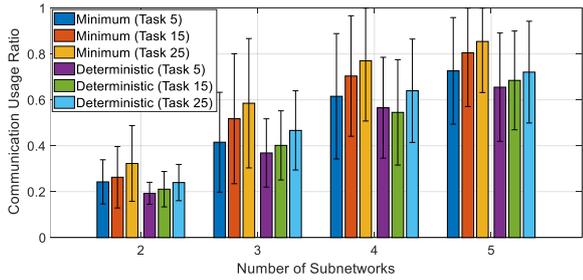

a) Communication resources utilization ratio of the selected link(s).

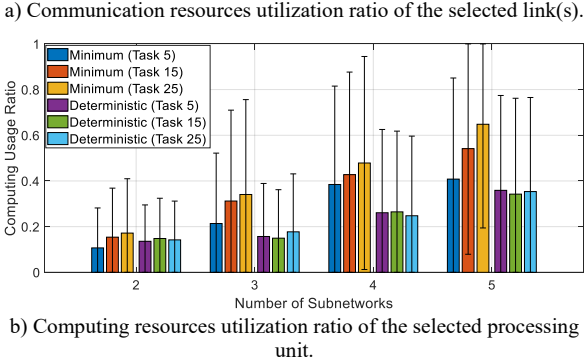

b) Computing resources utilization ratio of the selected processing unit.

Fig. 5. Average and standard deviation of resources utilization ratio when average SINR=30 dB (similar trends observed for average SINR = 0 dB).

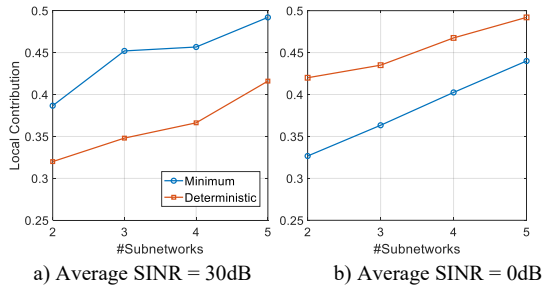

a) Average SINR = 30dB    b) Average SINR = 0dB

Fig. 6. Average ratio of tasks processed at subnetworks when the number of tasks per subnetwork is 15 (similar trends were observed for other values).

and computing resources (i.e. with a lower utilization ratio) compared to *Minimum*. For example, *Deterministic* selects communication links and processing units that are up to 17.5% (Fig. 5.a) and 45% (Fig. 5.b) less utilized – based on average values - than those selected by *Minimum*, respectively. The standard deviation values also show that *Deterministic* results in resource utilization ratios below 100% even in the scenarios with the largest number of subnetworks and task loads. In contrast, in scenarios with 4 and 5 subnetworks and 15 and 25 tasks per subnetwork, *Minimum* causes bottlenecks by reaching full resource utilization (100%), leading to system congestion, and potentially preventing certain tasks from meeting their deadlines.

By balancing the workload across the continuum (Fig. 5), *Deterministic* avoids putting excessive strain on the network's computing and communication resources, thus improving overall scalability. Fig. 5 shows that, on average, communication resources experience higher utilization than computing resources as the number of subnetworks and tasks increases. This is the case because all subnetworks share the same spectrum to connect to the wide-area cellular network and reach the edge node and cloud server. Fig. 6 shows the average ratio of tasks processed locally (within the subnetworks) by *Minimum* and *Deterministic*. Both schemes increase the ratio of tasks processed locally, i.e., within the subnetworks, as the number of subnetworks and total number of tasks grows. This is due to the scarcity of communication resources that all subnetworks must share to connect to the cellular network. However, Fig. 6 reveals that, compared to *Minimum*, *Deterministic* processes a higher proportion of tasks locally when the quality of the link from the subnetwork to the wide-area cellular network is poor (Fig. 6.b), whereas it relies less on local processing when the link quality is good (Fig. 6.a). These results reveal the ability of *Deterministic* to adapt the offloading of tasks based on the condition of shared communication resources, ensuring efficient and balanced task distribution across the continuum while augmenting the number of tasks that are executed before their deadline.

## V. CONCLUSIONS

This study has demonstrated that a deterministic approach to task offloading and (communication and computing) resource allocation in the IoT-edge-cloud continuum can enhance scalability in next-generation cellular networks. By flexibly managing task completion deadlines, a deterministic strategy achieves a more balanced workload and resource distribution across the continuum than existing methods and can better adapt to varying operating conditions (e.g., link quality). This augments task satisfaction ratios and fairness across the system while enabling more efficient resource utilization, which helps prevent resource saturation and enhances scalability.